\documentclass{emulateapj}

\newcommand{\subscript}[1]{\textnormal{\scriptsize{#1}}}
\newcommand{\rdout}{\ensuremath{R_\subscript{out}}}
\newcommand{\rdin}{\ensuremath{R_\subscript{in}}}

\newcommand{\tk}{\ensuremath{T_\subscript{K}}}

\newcommand{\irc}{IRC+10216}
\newcommand{\rstar}{\ensuremath{R_\star}}
\newcommand{\ethylene}{\ensuremath{\textnormal{C$_2$H$_4$}}}
\newcommand{\kms}{km~s$^{-1}$}
\newcommand{\cm}{cm$^{-1}$}

\newcommand{\science}{Sci}
\newcommand{\nature}{Natur}
\newcommand{\jms}{JMoSp}

\newcommand{\molphys}{MolPh}
\newcommand{\cjp}{CaJPh}

\newcommand{\ijs}{IJSp}
\newcommand{\jpb}{J. Phys. B: At. Mol. Opt. Phys.}

\begin{document}

\title{THE ABUNDANCE OF \ethylene{} IN THE CIRCUMSTELLAR ENVELOPE OF IRC+10216}
\shorttitle{Ethylene in IRC+10216}
\shortauthors{J. P. Fonfr\'{\i}a et al.}
\author{J. P. Fonfr\'ia\altaffilmark{1}}
\author{K. H. Hinkle\altaffilmark{2,4}}
\author{J. Cernicharo\altaffilmark{1}}
\author{M. J. Richter\altaffilmark{3,4}}
\author{M. Ag\'undez\altaffilmark{1}}
\author{L. Wallace\altaffilmark{2,5}}

\affiliation{$^1$Grupo de Astrof\'isica Molecular, Instituto de Ciencia de Materiales de Madrid,
CSIC, C/ Sor Juana In\'es de la Cruz, 3, Cantoblanco, 28049, Madrid (Spain)\\
$^2$National Optical Astronomy Observatory, P.O. Box 26732, Tucson, Arizona 
85726 (USA)\\
$^3$Physics Dept. - UC Davis, One Shields Ave., Davis,
CA 95616 (USA)}
\altaffiltext{4}{Visiting Astronomer at the Infrared Telescope Facility,
which is operated by the University of Hawaii under contract NNH14CK55B from 
the National Aeronautics and Space Administration.}
\altaffiltext{5}{Deceased}

\begin{abstract}
High spectral resolution mid-IR observations of ethylene (\ethylene{}) towards 
the AGB star \irc{} were obtained using the Texas Echelon Cross Echelle 
Spectrograph (TEXES) at the NASA Infrared Telescope Facility (IRTF). 
Eighty ro-vibrational lines from the 10.5~$\mu$m vibrational mode $\nu_7$ with 
$J\lesssim 30$ were detected in absorption. 
The observed lines are divided into two groups with rotational temperatures of 
105 and 400~K (warm and hot lines). 
The warm lines peak at $\simeq -14$~\kms{} with respect to the systemic 
velocity, suggesting that they are mostly formed outwards from 
$\simeq 20\rstar$. 
The hot lines are centered at $-10$~\kms{} indicating that they come from a 
shell between 10 and 20\rstar. 
35\% of the observed lines are unblended and can be fitted with a code 
developed to model the emission of a spherically symmetric circumstellar 
envelope. 
The analysis of several scenarios reveal that the \ethylene{} abundance 
relative to H$_2$ in the range $5-20$\rstar{} is $6.9\times 10^{-8}$ in average 
and it could be as high as $1.1\times 10^{-7}$. 
Beyond 20\rstar, it is $8.2\times 10^{-8}$. 
The total column density is $(6.5\pm 3.0)\times 10^{15}$~cm$^{-2}$. 
\ethylene{} is found to be rotationally under local thermodynamical equilibrium 
(LTE) and vibrationally out of LTE. 
One of the scenarios that best reproduce the observations suggests that up to 
25\% of the \ethylene{} molecules at 20\rstar{} could condense onto dust grains. 
This possible depletion would not influence significantly the gas acceleration 
although it could play a role in the surface chemistry on the dust grains.
\end{abstract}
\keywords{
line: identification --- 
line: profiles --- 
stars: AGB and post-AGB --- 
stars: carbon --- 
stars: individual (IRC+10216) ---
surveys
\vspace*{-0.5\baselineskip}
}

\maketitle

\section{Introduction}
\label{sec:introduction}
The planar asymmetric top molecule ethylene (\ethylene) is the simplest alkene
with a double bond linking the carbon atoms.
Due to its chemical reactivity, ethylene is one of the most important organic 
molecules expected to arise in a C-rich circumstellar environment 
\citep{millar_2000,woods_2003,cernicharo_2004,agundez_2006}.
Ethylene can be involved in the growth of hydrocarbons and in the formation of 
PAHs or dust grains \citep{contreras_2011}.

Despite the importance and reactivity displayed on Earth and its detection
in different objects of the solar system 
\citep*[e.g.,][]{encrenaz_1975,kim_1985,schulz_1999},
to date ethylene has been definitively detected beyond the solar system only 
in the circumstellar envelope (CSE) of the C-rich Asymptotic Giant Branch star 
(AGB) \irc{} based on the observation of several ro-vibrational lines of the 
$\nu_7$ band centered at 949~\cm{} 
\citep*[$\simeq 10.5~\mu$m;][]{betz_1981,goldhaber_1987,hinkle_2008}.
The scarcity of detections can be attributed to the lack of a permanent dipole 
moment, which results in the absence of a rotational spectrum in the mm 
wavelength range.
Hence, the only possible observation of ethylene is through its 
vibration-rotation lines in the mid-IR.

\irc{} is the most extensively studied AGB star due to its proximity 
\citep*[$\simeq 120$~pc;][]{groenewegen_2012}, and its chemical richness 
\citep*[e.g.,][]{cernicharo_1987,cernicharo_1989,cernicharo_2000,
cernicharo_2015b,hinkle_1988,bernath_1989,guelin_1991,keady_1993,kawaguchi_1995,
he_2008,patel_2011,agundez_2014}.
Surprisingly in this chemically complex circumstellar environment only a 
handful of the molecules detected are known to arise in the zone of gas 
acceleration between the stellar photosphere and $\simeq 20$~\rstar{}, i.e.
$\simeq 0\farcs4\simeq 50$~AU: CO, HCN, HNC, C$_2$H$_2$, CS, SiO, SiS, SiC$_2$, 
and Si$_2$C
\citep*[e.g.,][]{keady_1988,keady_1993,boyle_1994,cernicharo_1999,cernicharo_2010,
cernicharo_2013,cernicharo_2015b,fonfria_2008,fonfria_2014,fonfria_2015,decin_2010a,agundez_2012}. 
The gas acceleration process produces complex line profiles resulting from the 
combination of the molecular emission/absorption and the gas velocity field.
These profiles can be modeled to derive the molecular abundances as a function 
of the distance to the star.
In contrast, the radii where the spectral lines of some other molecules, for 
instance NH$_3$, SiH$_4$, H$_2$O, or \ethylene{}, are formed is still poorly 
known since they seem to arise after the gas has reached the terminal expansion 
velocity \citep*[$\gtrsim 20$~\rstar;][]{keady_1993,hinkle_2008,decin_2010b}.
This fact produces much simpler line profiles with little kinetic information.

In this paper, we analyze high-resolution mid-IR spectra of \irc{} in the 
spectral range $10.45-10.60~\mu$m.
A preliminary discussion of the observations can be found in \citet{hinkle_2008}.
In the next section we discuss the observations and the line identifications. 
The modeling of circumstellar envelope and the spectroscopic data used during
the fitting procedure are included in Section~\ref{sec:modeling}.
The results derived from the fits are presented and discussed in 
Section~\ref{sec:results}.

\section{Observations}
\label{sec:observations}

The observations were carried out with the Texas Echelon Cross Echelle 
Spectrograph \citep*[TEXES;][]{lacy_2002} mounted on the 3 meter NASA Infrared 
Telescope Facility (IRTF) on April 7, 2007 and May 29, 2008.
TEXES was used in the high-resolution mode with a resolving power 
$R\simeq 100,000$.
\irc{} was nodded along the slit for sky subtraction.
A black body-sky difference spectrum was used to correct for the atmosphere.
The data were reduced with the standard TEXES pipeline.
We normalized the spectrum acquired with each setting by removing the baseline 
with a 5th order polynomial fit.
The wavenumber scale has been set by sky lines.  

The 2007 spectrum was observed from $\simeq 924$ to 959~\cm{} using 11 
different grating settings.  
The observations were planned so the wavenumber coverage overlapped at each 
setting by approximately 50\%.
Thus the entire spectrum was observed twice. 
Each exposure covers about 5~\cm{} and 8 echelle orders. 
More than 80 segments of spectra had to be assembled into the final spectrum.
All the segments were plotted and sections of the orders that were near the 
edges and did not flat field properly were removed.
The overlap of the spectra was not perfect so there are small gaps.
Due to the trimming of bad sections the S/N is not uniform across the entire 
region observed.  

\begin{figure*}[hbt!]
\centering
\includegraphics[width=\textwidth]{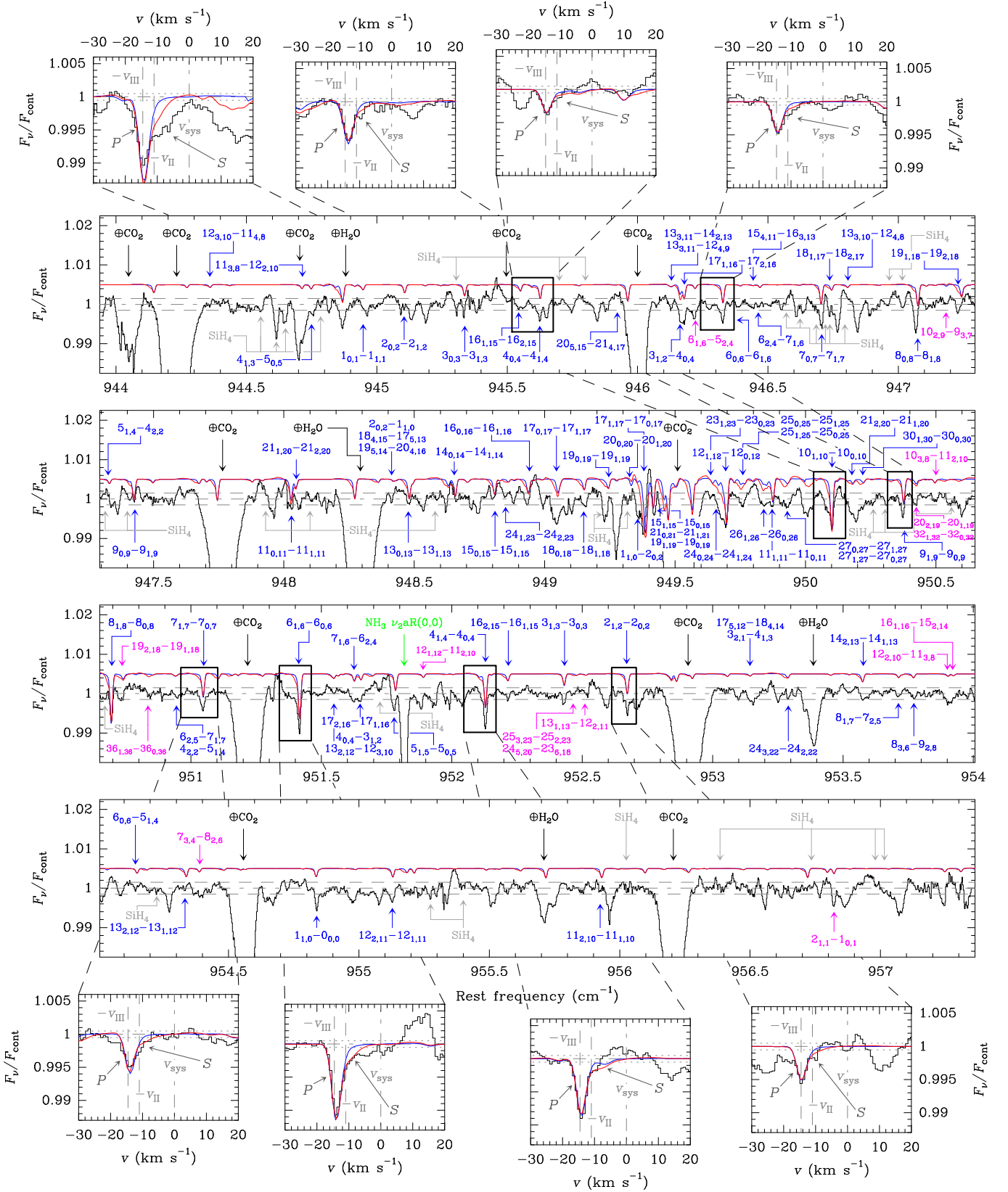}
\caption{Observed spectrum against rest frequency.
The spectral resolution is $\simeq 3$~\kms.
The gray dashed-dot and dashed horizontal lines indicate the continuum and
the detection limit ($3\sigma\simeq 0.15\%$).
The labels of the telluric lines are plotted in black.
The lines of \ethylene{} and NH$_3$ produced in the envelope of \irc{} 
undoubtedly identified are plotted in blue and green, respectively.
The features labeled with magenta comprises \ethylene{} lines that could be
fully blended with other unknown lines or are tentative identifications.
The spectrum also contains a number of SiH$_4$ lines plotted in gray.
The synthetic spectrum of \ethylene{} is plotted in blue (\ethylene{} only in 
Region II; Scenario 1, see Table~\ref{tab:table2}) and red (\ethylene{} in 
Regions II and III; Scenario $2A$).
The $P$ and $S$ labels indicate the primary and secondary absorptions 
(see text).
}
\label{fig:f1}
\end{figure*}

Examination of the spectra showed that the ethylene lines are indeed very weak
as reported more than 30 years ago by \citet{betz_1981} and a few years later 
by \citet{goldhaber_1987}. 
Typical depths of the strongest lines are less than 1\% of the continuum.
To improve the S/N, additional time was obtained in the 2008 observing season.
This data were obtained in the same way as 2007 but only covered the 944 to 
959~\cm{} region.
The signal-to-noise of the 2007 and 2008 spectra was similar so the two years 
were combined into the final observed spectrum.
We estimate the RMS of the random noise of the spectrum to be $\simeq 0.05\%$ of 
the continuum emission. 

The observed spectrum is not corrected for telluric contamination, which 
accounts for the strongest and prominent lines of CO$_2$ and H$_2$O 
(Fig.~\ref{fig:f1}).
Although weaker telluric lines are also present blocking some features coming 
from \irc, 80 \ethylene{} lines of the fundamental band $\nu_7$ with $J\le 30$ 
have been detected.
About 35\% of them are not blended and can be properly analyzed.
Although several lines of the fundamental bands $\nu_4$ and $\nu_{10}$ fall 
within the observed spectral range, none of the lines of these bands is above 
the detection limit ($3\sigma\simeq 0.15\%$ of the continuum).
No lines of hot or combination bands of the normal mode $\nu_7$ (e.g., 
$2\nu_7-\nu_7$ and $\nu_7+\nu_{10}-\nu_{10}$) have been found above the detection 
limit in this spectral range, either.
These lines involve ro-vibrational levels with energies $\gtrsim 1200$~K
\citep*[e.g.][]{oomens_1996},
barely populated in the mid and outer envelope where \ethylene{} is expected to 
exist \citep{cherchneff_1992}.
In addition to \ethylene{} and telluric lines, we were able to identify the 
$\nu_2$ aR(0,0) line of NH$_3$ at 951.8~\cm{} and a number of lines from 
SiH$_4$ between 944 and 947~\cm{} \citep{gray_1977,pierre_1986}.

We used the low spectral resolution \textit{ISO}/SWS observations carried out 
on 1996 May 31 \citep{cernicharo_1999} to determine the properties of the dusty 
envelope.
The uncertainty of these observations due to noise and the calibration process 
is estimated to be $\simeq 10\%$.

\section{\ethylene{} spectroscopic constants}
\label{sec:linelist}

The rest frequencies of the observed ethylene lines and their intensities were
taken from the HITRAN Database 2012 \citep{rothman_2013}.
\ethylene{} is an asymmetric top molecule with 12 vibrational modes.
Only the infrared active fundamental bands $\nu_{10}$, $\nu_7$, and $\nu_4$, 
with energies of 825.9, 948.8, and 1025.6~\cm, have lines in the observed range 
\citep*[e.g.,][]{ulenikov_2013}.
The $\nu_7$ band detected in our spectra corresponds to the out-of-plane 
bending mode. 

The partition function was computed by direct summation over all the
rotational levels with $J\lesssim 40$ of a number of vibrational states.
The energy of the vibrational states and the rotational constants involved in 
the calculation of the partition function ($E_\subscript{vib}\lesssim 5200$~K)
were retrieved from the HITRAN Database 2012, 
\citet{oomens_1996},
\citet{sartakov_1997},
\citet{willaert_2006}, 
\citet{loronogonzalez_2010},
\citet{lafferty_2011},
\citet{lebron_2013a},
\citet{lebron_2013b},
\citet{lebron_2013c}, and
\citet{ulenikov_2013}.
The energy of the rotational levels for each vibrational state was calculated 
by diagonalizing the Watson's Hamiltonian in its A-reduction $I^r$ 
representation, suitable to describe the \ethylene{} rotational structure
\citep*[e.g.,][]{lebron_2013a,ulenikov_2013}.
The number of vibrational states required to calculate the partition function
in warm regions of the envelope is less than 5 while up to some tens were 
necessary in hotter regions.
The partition function calculated under local thermodynamical equilibrium (LTE) 
and a kinetic temperature of 296~K is $1.09\times 10^4$, in very good agreement 
with previous calculations \citep{blass_2001,rothman_2013}.

The optical constants needed to calculate the opacity of dust grains were
taken from \citet{rouleau_1991} for amorphous carbon and from 
\citet{mutschke_1999} for SiC.

\section{Modeling}
\label{sec:modeling}

The modeling process was done with the code developed by \citet{fonfria_2008}, 
which numerically solves the radiation transfer equation of a spherically 
symmetric envelope composed of gas and dust in radial expansion.
It was successfully used to analyzed the mid-IR spectra of C$_2$H$_2$, HCN, 
SiS, C$_4$H$_2$, and C$_6$H$_2$ at 8 and $13~\mu$m towards \irc{} and the 
proto-planetary nebula CRL618 \citep{fonfria_2008,fonfria_2011,fonfria_2015}.
A detailed description of the code can be found in \citet{fonfria_2008} and
\citet{fonfria_2014}.

In order to analyze the \ethylene{} lines identified in the spectrum, we 
adopted the envelope structure and expansion velocity field derived by 
\citet{fonfria_2015} from the modeling of the SiS emission.
The envelope is composed of three Regions (I, II, and III) ranging from 1 to 
5\rstar, from 5 to 20\rstar, and from 20\rstar{} up to the external radius of 
the envelope, $R_\subscript{ext}$.
This parameter was fixed to 500\rstar, a distance compatible with the position 
where \ethylene{} is believed to be dissociated
\citep*[e.g.,][]{glassgold_1996,guelin_1997,agundez_2006}.
It is also large enough to prevent the uncertainties of the synthetic 
absorption components of the lines and of the calculated continuum emission to 
be above the noise RMS of our data.
The gas expansion velocity field, $v_\subscript{exp}$, comprises a linear 
increment between 1 and 11~\kms{} along Region I and two zones matching up with 
Regions II and III where the gas expands at a constant velocity of 11 and 
14.5~\kms, respectively (Table~\ref{tab:table1}).
The adopted line width is assumed to be 5~\kms{} at the stellar photosphere, 
following the power-law $\propto 1/r$ up to 5\rstar{} and remaining equal to 
1~\kms{} further away \citep{agundez_2012}.
The gas density and rotational and vibrational temperature profiles are assumed 
to follow the laws $\propto r^{-2} v_\subscript{exp}^{-1}$ and $r^{-\gamma}$, 
respectively, where $\gamma$ could be different for the rotational and 
vibrational temperatures.
No dust has been considered to exist in Region I.
Throughout the rest of the envelope we have adopted dust grains composed of 
amorphous carbon (AC) and SiC with a size of $0.1~\mu$m and a density of 
2.5~g~cm$^{-3}$.

\begin{deluxetable}{cccc}
\tabletypesize{\footnotesize}
\tablecolumns{4}
\tablewidth{0.475\textwidth}
\tablecaption{Non-free parameters in the fits of lines of \ethylene\label{tab:table1}}
\tablehead{\colhead{Parameter} & \colhead{Units} & \colhead{Value} & \colhead{Ref.}}
\startdata
$D$                               & pc                & 123                & 1 \\
$\alpha_\star$                     & arcsec            & 0.02               & 2 \\
\rstar                            & cm                & $3.7\times 10^{13}$ & \\
$\dot M$                          & M$_\odot$~yr$^{-1}$ & $2.1\times 10^{-5}$ & 3 \\
$T_\star$                          & K                 & $2330$             & 2 \\
$\rdin$                           & \rstar            & $5$                & 3 \\
$\rdout$                          & \rstar            & $20$               & 3 \\
$v_{\subscript{exp}}(1\rstar\le r<\rdin)$ & \kms         & $1+2.5(r/\rstar-1)$ & 4 \\
$v_\subscript{exp}(\rdin\le r<\rdout)$   & \kms         & $11.0$             & 3 \\
$v_\subscript{exp}(r\ge\rdout)$         & \kms          & $14.5$             & 3 \\
$\tk(\rstar\le r< 9\rstar)$ & K             & $T_\star(\rstar/r)^{0.58}$  & 5   \\
$\tk(9\rstar\le r< 65\rstar)$ & K           & $\tk(9\rstar)(9\rstar/r)^{0.40}$ & 5   \\
$\tk(r\ge 65\rstar)$    & K                 & $\tk(65\rstar)(65\rstar/r)^{1.2}$ & 5   \\
$\Delta v(1\rstar\le r<\rdin)$    & \kms              & $5\left(\rstar/r\right)$                & 6 \\
$\Delta v(r\ge\rdin)$             & \kms              & $1$                & 6 \\
$f_\subscript{AC}$                  & \%                & 95                 & 3 \\
$f_\subscript{SiC}$                 & \%                & 5                  & 3 \\
$\tau_\subscript{dust}(11~\mu$m)    &                   & 0.70               & 4 \\
$T_\subscript{dust}(\rdin)$         & K                 & 825                & 4 \\
$\gamma_\subscript{dust}$           &                   & 0.39               & 4 
\enddata
\tablecomments{
$D$: distance to the star; 
$\alpha_\star$: angular stellar radius;
$R_\star$: linear stellar radius;
$\dot M$: mass-loss rate; 
$T_\star$: stellar effective temperature;
\rdin{} and \rdout: position of the outer boundaries of Regions I and II;
$v_\subscript{exp}$: gas expansion velocity;
$\Delta v$: line width;
$f_\subscript{X}$: fraction of dust grains composed of material X;
$\tau_\subscript{dust}$: dust optical depth along the line-of-sight;
$T_\subscript{dust}$: temperature of dust grains;
$\gamma_\subscript{dust}$: exponent of the decreasing dust temperature power-law
($\propto r^{-\gamma_\subscript{dust}}$).
References: 
(1) \citet{groenewegen_2012}
(2) \citet{ridgway_1988}
(3) \citet{fonfria_2008}
(4) \citet{fonfria_2015}
(5) \citet{debeck_2012}
(6) \citet{agundez_2012}.
}
\end{deluxetable}

\subsection{Fitting procedure}
\label{sec:fitting.procedure}

The fitting procedure is based on the minimization of the $\chi^2$ function 
defined as
\begin{equation}
\chi^2=\frac{1}{n-p}\sum_{i=1}^n w_i \left[
\left(\frac{F_\nu}{F_\subscript{cont}}\right)_\subscript{obs,$i$}-
\left(\frac{F_\nu}{F_\subscript{cont}}\right)_\subscript{synth,$i$}\right]^2,
\label{eq:chi2}
\end{equation}
where $n$ and $p$ are the number of frequency channels and free parameters
involved in the fitting process, $(F_\nu/F_\subscript{cont})_\subscript{synth}$ and
$(F_\nu/F_\subscript{cont})_\subscript{obs}$ are the flux normalized to the 
continuum for the synthetic and observed spectra, and $w$ is a weight included 
for convenience (see Section~\ref{sec:weight}).
We assumed a number of physical parameters related to the envelope model as 
fixed during the fitting process (gas expansion velocity field, H$_2$ density, 
kinetic temperature, and terminal line width, among others; see 
Tables~\ref{tab:table1} and \ref{tab:table2}).
The physical and chemical quantities related to \ethylene{} derived from the 
fits are the abundance with respect to H$_2$, the rotational temperature at 
20\rstar, and the exponent of the power-law followed by the rotational 
temperature beyond 20\rstar.
All of the free parameters have been varied until the whole set of lines has 
been best fitted at the same time.
The uncertainties of the free parameters have been estimated by simultaneously
varying all of them until the synthetic spectrum departs at any frequency in 
more than the observational errors from the best global fit.

\begin{deluxetable*}{cccccccc}
\tabletypesize{\footnotesize}
\tablecolumns{8}
\tablecaption{Derived Parameters\label{tab:table2}}
\tablehead{\colhead{Parameter} & \colhead{Units} & \colhead{Value}  & \colhead{Error} & \colhead{Value} & \colhead{Error} & \colhead{Value} & \colhead{Error}}
\startdata
 & & \multicolumn{2}{c}{\textit{Scenario 1}} & \multicolumn{4}{c}{\textit{Scenario 2}} \\
 & & & & \multicolumn{2}{c}{\textit{Case A}} & \multicolumn{2}{c}{\textit{Case B}} \\
\hline
$\chi_d^2$                 & $\times 10^{-2}$ & 4.97                 & ---         & 3.84              & ---         & 3.92              & --- \\
$\chi_\subscript{red}^2$     &                 & 7.02                 & ---         & 3.80               & ---         &4.08               & --- \\
$x(\rstar\le r\le\rdin^-)$ & $\times 10^{-8}$ & 0.0                  & ---         & 0.0                & ---         &0.0                & --- \\
$x(r=\rdin^+)$             & $\times 10^{-8}$ & 0.0                  & ---         & 6.9                & $+1.9/-1.8$ &0.0                & --- \\
$x(r=\rdout^-)$            & $\times 10^{-8}$ & 0.0                  & ---         &                    &             &11.0               & $+3.3/-2.6$ \\
$x(r\ge\rdout^+)$          & $\times 10^{-8}$ & 8.2                  & $+1.1/-1.2$ & 8.2                & $+1.3/-1.1$ &8.2                & $+1.3/-1.0$ \\
$T_\subscript{rot}(\rstar)$  & K               & 2330$^*$             & ---         & 2330$^*$           & ---          &2330$^*$           & --- \\
$T_\subscript{rot}(\rdin)$   & K               & 916$^*$              & ---         & 916$^*$            & ---          &916$^*$            & --- \\
$T_\subscript{rot}(\rdout)$  & K               & 370                  & $+60/-80$   & 380                & $+65/-60$   &380                & $\pm 60$ \\
$\gamma_\subscript{rot}$     &                 & 0.54                 & $\pm 0.04$  & 0.54$^\dagger$      & ---         &0.54$^\dagger$      & --- \\
$T_\subscript{vib}(\rstar)$  & K               & 2330$^\ddagger$       & ---         & 2330$^\ddagger$      & ---         & 2330$^\ddagger$     & --- \\
$T_\subscript{vib}(\rdin)$   & K               & 500$^\ddagger$        & ---         & 500$^\ddagger$       & ---         & 500$^\ddagger$      & --- \\
$T_\subscript{vib}(\rdout)$  & K               & 315$^\ddagger$        & ---         & 315$^\ddagger$       & ---         & 315$^\ddagger$      & --- \\
$\gamma_\subscript{vib}$     &                 & 1.0$^\ddagger$        & ---         & 1.0$^\ddagger$       & ---         & 1.0$^\ddagger$      & ---\\
$N_\subscript{col}(\rdin^+\le r\le\rdout^-)$  & $\times 10^{15}$~cm$^{-2}$  & 0.0   & ---            &  5.8   & $+1.6/-1.5$    & 2.5  & $+0.8/-0.6$ \\
$N_\subscript{col}(r\ge\rdout^+)$             & $\times 10^{15}$~cm$^{-2}$  & 1.66  & $+0.22/-0.24$  &  1.67  & $+0.26/-0.22$  & 1.67  & $+0.26/-0.20$ \\
$N_\subscript{col,total}$                     & $\times 10^{15}$~cm$^{-2}$  & 1.66  & $+0.22/-0.24$  &  7.5   & $+1.9/-1.7$    & 4.2  & $+1.1/-0.8$
\enddata
\tablecomments{
$\chi_d^2$ and $\chi_\subscript{red}^2$: minimum of the $\chi^2$ function 
(Eq.~\ref{eq:chi2}) assuming the weights $w=1/d^2$ and $1/\sigma^2$, where $d$ 
is the depth of each line and $\sigma$ is the noise RMS of the spectrum 
($\chi_\subscript{red}^2$ is the reduced $\chi^2$ function);
$x$: abundance relative to H$_2$;
$\rdin^+$ and $\rdin^-$: \rdin{} approached from Region II and Region I, 
respectively;
$\rdout^+$ and $\rdout^-$: \rdout{} approached from Region III and Region II, 
respectively;
$T_\subscript{rot}$: rotational temperature;
$\gamma_\subscript{rot}$: exponent of the rotational temperature power-law 
outwards from $\rdout^+$ ($\propto r^{-\gamma_\subscript{rot}}$);
$T_\subscript{vib}$: vibrational temperature derived from band $\nu_7$;
$\gamma_\subscript{vib}$: exponent of the vibrational temperature power-law 
outwards from \rdout{} ($\propto r^{-\gamma_\subscript{vib}}$);
$N_\subscript{col}$: \ethylene{} column density calculated with the parameters
derived from the best fits to the observations.
The parameters for which no uncertainty has been provided (---) were assumed
as fixed during the whole fitting process.
The errors are 1-sigma uncertainties for all the parameters.
$^*$This rotational temperature was fixed to the kinetic temperature derived by 
\citet{debeck_2012}.
$^\dagger$This value was fixed to the result derived from the fit to the observed
lines assuming Scenario 1.
$^\ddagger$The vibrational temperature for \ethylene{} was fixed to the 
vibrational temperature derived by \citet{fonfria_2008} from the analysis 
of the C$_2$H$_2$ lines of band $\nu_5$.
A blank cell contains the same value than the immediately upper cell.
}
\end{deluxetable*}

\subsection{Weighting the $\chi^2$ function}
\label{sec:weight}

The strongest lines identified in the spectrum are those with $J\lesssim 15$
(Fig.~\ref{fig:f1}).
The formation of \ethylene{} in the different regions of the envelope could be 
influenced by variations in the physical and chemical conditions between the 
stellar photosphere and $\simeq 50$\rstar{} 
\citep{fonfria_2014,fonfria_2015}, probably related to the binary nature of
the central stellar system 
\citep{cernicharo_2015a,cernicharo_2016,decin_2015,kim_2015,quintana-lacaci_2016}.
This means that the difference between the strongest observed and synthetic 
lines could be as large as the depth of the absorption component of the weakest
lines.
This fact could result in an underestimation of the importance of the weakest 
lines (usually the highest excitation lines) during the automatic fitting 
procedure.
Thus, the rotational temperature of the \ethylene{} enclosed by the inner 
shells of Region III would be lower than the actual one, being dominated by the 
temperature prevailing in the outer envelope.

This situation can be solved by adopting the weight $w=1/d^2$ for all the 
channels of each line, where $d$ is the depth of its absorption component.
This choice gives a comparable importance during the fitting process to the 
weak lines formed at distances to the star $\lesssim 30\rstar$ and the strong 
lines mostly formed beyond, resulting in a more realistic rotational 
temperature profile. 
It is worth noting that the fit to the strong lines and the derived abundance 
profile throughout Region III are not significantly affected by the chosen 
weight.

It is also convenient to use the so-called \textit{reduced $\chi^2$ function}, 
$\chi^2_\subscript{red}$, to numerically estimate the goodness of the fits.
Its definition involves a weight $w=1/\sigma^2$, where $\sigma$ is the 
uncertainty of the observed data, and the number of degrees of freedom of the
fit, usually defined as $n-p$, where $n$ is the number of points of the set to 
fit and $p$ the number of parameters varied during the fitting process.
In theory, $\chi_\subscript{red}^2$ approaches to 1 for a good fit to a set of 
experimental data, adopting larger values for worse fits.
In practice, it is difficult to use $\chi_\subscript{red}^2$ to figure out the 
actual quality of a given fit but it can be used to compare different fits 
\citep*[e.g.,][]{taylor_1997,tallon-bosc_2007}.

\newpage

\subsection{Deviations of the synthetic spectrum from the observed one}

The fit to the observed spectrum was improved by slightly shifting the 
frequency of several observed \ethylene{} lines.
All of these shifts are smaller than 2.5~\kms{} ($\simeq 0.007$~\cm)
depending on the chosen model (see Section~\ref{sec:abundance}).
These differences can be explained by the spectral resolution of our data, 
estimated to be $\simeq 3$~\kms.
However, the deviation between the synthetic and observed profiles of some 
lines such as, e.g., $1_{1,0}-0_{0,0}$ and $8_{1,8}-8_{0,8}$ (2 and 3.5 times 
smaller and larger, respectively), are unexpectedly large while other lines 
involving ro-vibrational states with similar or intermediate energies are well 
reproduced (e.g., $2_{1,2}-2_{0,2}$, $6_{0,6}-6_{1,6}$, $9_{1,9}-9_{0,9}$).
The observed feature identified as line $1_{1,0}-0_{0,0}$ could be stronger than 
expected due to an unnoticed blending with other unknown lines or a cold 
contribution not included in our model.
Nevertheless, the fact that the observed $8_{1,8}-8_{0,8}$ line is weaker than 
the synthetic one while other lines formed in the same region of the envelope 
are better reproduced suggests that, probably, the baseline was not well 
removed around this line.
This problem usually appears when the number of molecular features in the 
vicinity of the target line is high, hiding the actual baseline.

\vspace*{2\baselineskip}

\section{Results and discussion}
\label{sec:results}

About 35\% of the detected lines have well enough defined profiles to be fitted.
Most lines show an absorption component.
A systematic emission component is not obvious in the data. 
Some lines showing a possible emission component are probably affected by 
the baseline removal.
The absorption component of the strong lines is usually dominated by a deep, 
narrow contribution at $\simeq -14$~\kms{} (primary absorption, $P$ in the 
insets of Fig.~\ref{fig:f1}), presumably caused by the gas expanding at the 
terminal velocity, and a red-shifted weaker contribution (secondary absorption, 
$S$ in the insets of Fig.~\ref{fig:f1}).
These secondary absorptions could be argued to be other lines of the same band 
or of a hot band.
In fact, some strong lines of the $\nu_7$ band are accompanied by close, weaker 
lines of the same band that influence their profiles.
However, these \ethylene{} line sets are not so common and the frequency shift 
between the primary and the secondary absorptions remains nearly constant for 
every observed strong line.
This would be an unexpected behavior if the secondary absorptions were other 
\ethylene{} lines.
The Doppler velocity of the secondary absorptions is more compatible with the 
features produced by the gas in Region II expanding at $\simeq 11$~\kms.
This reasoning is supported as well by the existence of weak lines involving 
ro-vibrational levels with energies up to $\simeq 800$~\cm{} ($\simeq 1150$~K) 
that are expected to be significantly excited inwards from $\simeq 20$\rstar, 
where the kinetic temperature is $\gtrsim 470$~K \citep{debeck_2012}.
On the other hand, the lack of an evident emission component in the observed 
lines and, particularly, in lines involving lower ro-vibrational levels with 
energies $\gtrsim 1200$~K suggest that the \ethylene{} abundance is negligible 
from the stellar surface up to the middle shells of Region II 
($\simeq 10-15$\rstar).
This spatial distribution is roughly consistent with the predicted by 
thermodynamical chemical equilibrium models \citep*[e.g.,][]{cherchneff_1992},
which suggest that \ethylene{} reaches a significant abundance in shells with a 
kinetic temperature $\gtrsim 1000$~K (closer to the star than 5\rstar{} in our 
envelope model).

\begin{figure}
\centering
\includegraphics[width=0.475\textwidth]{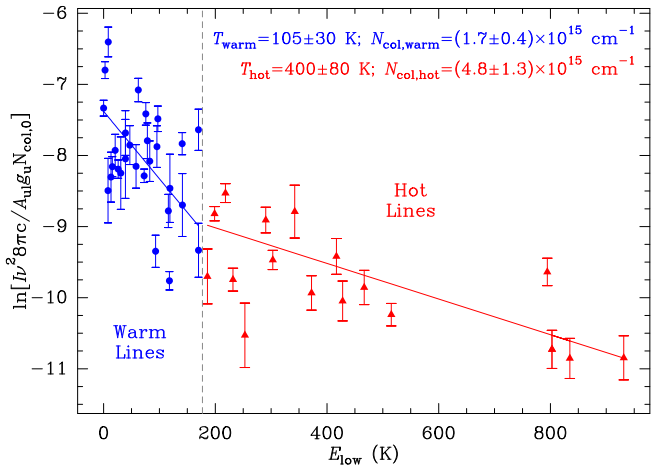}
\caption{Ro-vibrational diagram of the observed \ethylene{} lines.
The lines are separated into two different groups: warm lines (blue 
dots) and hot lines (red triangles).
The warm lines are those involving lower ro-vibrational levels with energies 
below 177~K.
The hot lines involve lower levels with higher energies.
These groups of lines were linearly fitted separately following 
Eq.~\ref{eq:boltzmann} to derive their rotational temperatures, $T_\subscript{rot}$, 
and their column densities, $N_\subscript{col}$.
}
\label{fig:f2}
\end{figure}

A quantitative analysis can be obtained through a ro-vibrational diagram (see 
Fig.~\ref{fig:f2}).
In this plot, we take advantage of the absence of an emission component of the 
lines.  
When such emission is present, it overlaps the absorption component of the 
profile requiring the application of complex envelope models to derive reliable 
information about the circumstellar gas 
\citep*[e.g.,][]{keady_1988,keady_1993,boyle_1994,fonfria_2008,fonfria_2015}.
The general weakness of the observed \ethylene{} lines, with depths of their
absorption components below $\simeq 1\%$ of the continuum, suggest that they
are all optically thin and the following formula holds:
\begin{equation}
\label{eq:boltzmann}
\ln{\left[\frac{I\nu^28\pi c}{A_{ul}g_uN_\subscript{col,0}}\right]}\simeq
\ln{\left[\frac{N_\subscript{col}}{N_\subscript{col,0}Z}\frac{\theta_s^2}{\theta_s^2+\theta_b^2}\right]}
-\frac{hcE_\subscript{low}}{k_\subscript{B}T_\subscript{rot}}
\end{equation}
where $\nu$ is the rest frequency of each line (\cm), $I$ is the integral of 
the normalized flux absorbed by each line over the frequency 
($\int_\subscript{line}\left(1-F_\nu/F_\subscript{cont}\right)d\nu$; \cm), 
$A_{ul}$ the A-Einstein coefficient of the transition (s$^{-1}$), $g_u$ the 
degeneracy of the upper level of the transition, $Z$ is the partition function, 
$c$ the velocity of light in vacuum (cm~s$^{-1}$), $N_\subscript{col}$ the column 
density of \ethylene{} (cm$^{-2}$), $E_\subscript{low}$ the energy of the lower 
ro-vibrational level involved in the transition (\cm), $h$ and 
$k_\subscript{\tiny B}$ are the Planck and Boltzmann constants, and 
$T_\subscript{rot}$ the rotational temperature (K).
$N_\subscript{col,0}=10^{15}$~cm$^{-2}$ is a fixed column density included for 
convenience to get a dimensionless argument for the logarithms.
In this equation, we assume the rotational temperature is low enough that 
$e^{-hc\nu/k_\subscript{\tiny B} T_\subscript{\tiny rot}}\ll 1$.
The factor $\theta_s^2/(\theta_s^2+\theta_b^2)$, where $\theta_s$ is the angular 
size of the source and $\theta_b$ the HPBW of the IRTF ($\simeq 0\farcs9$ at 
10.5~$\mu$m), accounts for the effect of the size differences between the 
source and the main beam of the PSF.
The size of the source can be roughly estimated with the aid of our code
assuming the parameters of Table~\ref{tab:table1} and that ethylene is under 
LTE.
This approximation gives that $\theta_s\simeq 1\farcs9\pm 0\farcs2$, 
which means that $\theta_s^2/(\theta_s^2+\theta_b^2)\simeq 0.82\pm 0.03$.

The data plotted in Fig.~\ref{fig:f2} suggest that the observed lines belong to 
two different populations.
We will refer to them as warm and hot lines, hereafter.
The data related to each population were separately fitted adopting the typical
weight based on the statistical uncertainties of the line intensities.
The warm lines involve lower ro-vibrational levels with energies below 
$\simeq 180$~K and show a rotational temperature of $105\pm 30$~K,
compatible with the preliminary result proposed by \citet{hinkle_2008}. 
Adopting the envelope model described in Section~\ref{sec:modeling} and LTE, 
these lines would be mostly formed at distances to the star of 
$\simeq 125-200\rstar$ ($\simeq 2\farcs5-4\farcs0$).
On the other hand, the hot lines are formed in shells where the physical 
conditions favor a higher rotational temperature ($T_\subscript{rot}=400\pm 80$~K).
Thus, these shells would be located at $\simeq 20-50\rstar$ 
($\simeq 0\farcs4-1\farcs0$) from the star.
This means that the existence of a significant abundance of \ethylene{} in 
Region I, i.e., between the stellar photosphere and 5\rstar, can be ruled out.
The total column density derived from the ro-vibrational diagram is
$(6.5\pm 1.7)\times 10^{15}$~cm$^{-2}$ compatible with the rough estimations by 
\citet{betz_1981} and \citet{goldhaber_1987}, i.e., $\sim 10^{16}$ and 
$4\times 10^{15}$~cm$^{-2}$, respectively, which are affected by an error of one 
order of magnitude.
About 75\% of the \ethylene{} column density is enclosed by the shell at 
20\rstar{} while only $\simeq 25\%$ of it is related to the ethylene spread at 
larger distances.

\subsection{Abundance with respect to H$_2$}
\label{sec:abundance}

In order to describe the \ethylene{} abundance profile we attempted a synthesis
of the observed lines assuming two different scenarios:
(1) \ethylene{} only exists in Region III and
(2) \ethylene{} exists in Regions II and III.
Scenario 2 is further broken down into two sub-scenarios, $A$ and $B$, where we 
assumed that the abundance is constant throughout Regions II and III (Scenario 
$2A$) and it is constant in Region III, growing linearly from 0 across Region II
up to an abundance that could be different than that of Region III (Scenario 
$2B$).
In both Scenarios $2A$ and $2B$ the abundance was not necessarily a continuous
profile.
The results of the analysis of all these scenarios can be found in
Table~\ref{tab:table2}.

The synthetic spectrum that fits the observations worst is from Scenario 1 
with $\chi_\subscript{red}^2\simeq 7.0$ (see Table~\ref{tab:table2}).
This scenario is capable of reproducing fairly well the primary absorption
of the observed lines but it fails to describe their secondary absorptions and
the high excitation lines.
Scenarios $2A$ and $2B$ explain better the shape of the strong lines by adding 
a secondary absorption, reproducing at the same time the high excitation lines.
The synthetic spectra calculated for both show an average 
$\chi_\subscript{red}^2\simeq 4.0$, significantly smaller than for Scenario 1.
None of these scenarios are able to fit the observations with 
$\chi_\subscript{red}^2\simeq 1.0$, which would mean a perfect correlation 
between the observations and the synthetic spectrum.
This fact indicates that the actual abundance distribution is likely more 
complex than we have assumed here.
The shape of the secondary absorption of the strongest lines, which are usually 
deeper and narrower than seen in the synthetic spectra, might indicate that 
the solid angle subtended by the absorbing region around 20\rstar{} is smaller 
than expected for a spherically symmetric abundance distribution.
Consequently, there is an absorption deficit close to the systemic velocity 
(0~\kms{} in the insets of Fig.~\ref{fig:f1}) in the observed lines with 
respect to the synthetic features.
This asymmetry in the abundance distribution could be related to those that
\citet{fonfria_2014} invoked to explain the high angular resolution 
interferometer observations of H$^{13}$CN, SiO, and SiC$_2$ at 1.2~mm or to the 
small scale features related to the spiral structure recently found in the 
envelope of \irc{} 
\citep{cernicharo_2015a,cernicharo_2016,quintana-lacaci_2016}.

\begin{figure}
\centering
\includegraphics[width=0.475\textwidth]{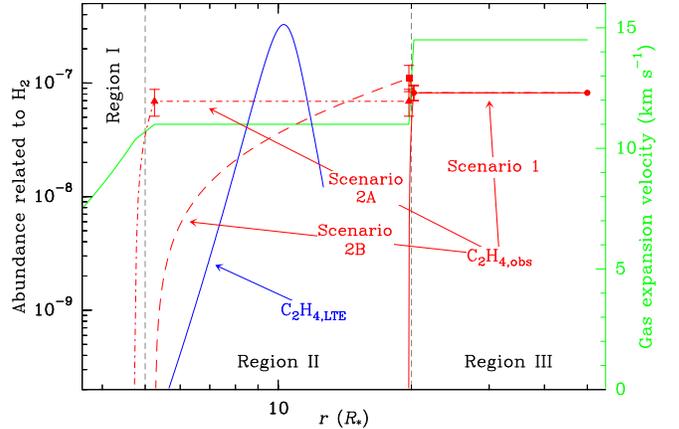}
\caption{Ethylene abundance with respect to H$_2$.
The profiles derived for Scenarios 1, $2A$, and $2B$ are plotted in red (solid, 
dash-dotted, and dashed, respectively).
A synthetic profile calculated with a chemical model under TE (see text) is 
plotted in blue.
The adopted gas expansion velocity profile is plotted in green.}
\label{fig:f3}
\end{figure}

The \ethylene{} abundance with respect to H$_2$ throughout Region III derived 
from our best fit is $(8.2\pm 1.3)\times 10^{-8}$ for all the Scenarios (1, 
$2A$, and $2B$; see Figure~\ref{fig:f3}), which means a column density of 
$(1.7\pm 0.3)\times 10^{15}$~cm$^{-2}$.
This column density is in very good agreement with the value derived from the 
linear fit to the warm lines in the ro-vibrational diagram 
[$(1.7\pm 0.4)\times 10^{15}$~cm$^{-2}$; Fig.~\ref{fig:f2}].
The \ethylene{} abundance profile in Region II can be roughly explained by the 
two Scenarios $2A$ and $2B$.
Although Scenario $2A$ best reproduces the observations with 
$\chi_\subscript{red}^2\simeq 3.8$, the synthetic spectrum calculated assuming 
Scenario $2B$ shows a $\chi_\subscript{red}^2$ only 7\% higher.
The abundance derived from Scenarios $2A$ and $2B$ in Region II are compatible.
However, Scenario $2B$ enhances the absorption of lines involving low energy
ro-vibrational levels formed in the outer shells of Region II at the expense of 
the absorption of higher excitation lines, formed closer to the star.
Scenario $2A$ results in stronger high excitation lines but the difference with
the same lines calculated with Scenario $2B$ is always below the noise RMS of 
the observed spectrum.
These facts suggest that both Scenarios are similarly possible and we are 
barely sensitive to the absorption produced in the inner half of Region II.
The column density in Region II derived from the hot lines of the
ro-vibrational diagram 
[$(4.8\pm 1.3)\times 10^{15}$~cm$^{-2}$; Fig.~\ref{fig:f2}] 
is between the values derived from Scenarios $2A$ and $2B$ [$(5.8\pm 1.6)$ and 
$(2.5\pm 0.8)\times 10^{15}$~cm$^{-2}$, respectively; Table~\ref{tab:table2}],
which indicates that the best scenario seems to be closer to Scenario $2A$, 
although it could also be a mix of both since the differences are not 
particularly significant.
Thus, the actual abundance could be constant in the outer half (or a larger 
fraction) of Region II, linearly increasing in the inner half.
In any case, the resulting spectrum would differ to those derived from 
Scenarios $2A$ and $2B$ in less than the noise RMS and would not throw any 
further light on the problem.

In a circumstellar envelope of an AGB star with a ratio C/O~$\simeq 2$, such as 
\irc{} \citep*[e.g.,][]{agundez_2006}, \ethylene{} is expected to form under 
thermal equilibrium (TE) near the stellar photosphere with a very low abundance 
with respect to H$_2$ \citep*[$\sim 10^{-15}$; e.g.,][]{cherchneff_1992}.
The abundance increases when the kinetic temperature decreases reaching a 
terminal value that keeps constant up to the outer envelope, where \ethylene{} 
is dissociated by the Galactic UV radiation.
The position where the terminal abundance is reached and its value are hard to 
calculate since the thermal equilibrium condition seems to hold only from the 
stellar photosphere up to $\simeq 3-7\rstar$ 
\citep{cernicharo_2010,cernicharo_2013,fonfria_2014}.
Kinetic chemistry models required to make predictions at larger distances from 
the star are substantially affected by the uncertainties related to the kinetic 
constants involved in the chemical reaction network and the complexity of the 
calculations.

\begin{deluxetable}{cccc}
\tabletypesize{\footnotesize}
\tablecolumns{4}
\tablewidth{0.475\textwidth}
\tablecaption{Previous estimates of the terminal \ethylene{} abundance/column density\label{tab:table3}}
\tablehead{\colhead{\parbox[t]{0.15\textwidth}{\centering $x$\\[0.3\baselineskip] ($\times 10^{-8}$)}} & \colhead{\parbox[t]{0.225\textwidth}{\centering $N_\subscript{col,total}$\\($\times 10^{15}$~cm$^{-2}$)}} & \colhead{Method} & \colhead{Reference}}
\startdata
$5.5^*-8.2$ & $4.2-7.5$ & Obs. & 1\\
10 & $10-100$ & Obs. & 2\\
$1-2^\dagger$ & $4-40$ & Obs. & 3\\
30$^\ddagger$ & --- & Model & 4\\
0.1 & --- & Model & 5\\
0.2 & 0.001 & Model & 6
\enddata
\tablecomments{
$x$: \textit{maximum terminal} \ethylene{} abundance with respect to H$_2$;
$N_\subscript{col,total}$: total \ethylene{} column density;
Method: derived from observations (Obs.) or from a chemical model (Model);
References: 
(1) This work (beyond 20\rstar)
(2) \citet{betz_1981}
(3) \citet{goldhaber_1987}
(4) \citet{cherchneff_1992}
(5) \citet{doty_1998}
(6) \citet{millar_2000}.
$^*$This value is the average abundance of Scenario $2B$.
$^\dagger$This value has been calculated with the aid of the CO abundances 
with respect to H$_2$ adopted by \citet{keady_1988} and \citet{debeck_2012}.
$^\ddagger$This abundance corresponds to a C-rich AGB star with 
C/O~$\simeq 1.5-3.0$.
}
\end{deluxetable}

There are only a few estimates for the \ethylene{} column density or the 
abundance with respect to H$_2$ derived from observations or calculated using 
chemical models for \irc{} in the literature to compare with our results (see 
Table~\ref{tab:table3}).
The lower limit of the column density derived by \citet{betz_1981} and 
\citet{goldhaber_1987} are compatible with our results within a factor of 2.
These authors considered an error of one order of magnitude to account for
the effect of the dust emission on the \ethylene{} lines.
However, this uncertainty is too big because the optical depth of the dust is 
$\simeq 0.75$ at $\simeq 10.5~\mu$m (Table~\ref{tab:table1}).
In fact, this effect exists but it is much less important than they considered.
The chemical model by \citet{millar_2000} gives a very low column density
in part because their model produces ethylene in the outer envelope, where the
gas density is orders of magnitude lower than at in the inner layers of the CSE.
This is the same reason that the abundance with respect to H$_2$ predicted
in \citet{millar_2000} and \citet{doty_1998} are about two orders of magnitude
lower than ours.
The peak abundance calculated by \citet{cherchneff_1992} for a C-rich AGB star 
with C/O~$\simeq 1.5-3.0$ and assuming TE is a factor of 4 to 6 higher than 
our estimate.
This disagreement could be related to the fact that their envelope model
represents a typical AGB star different than \irc.
However, our abundance is in good agreement with that proposed by 
\citet{betz_1981}, although these authors only estimated the order of magnitude.
Interestingly, despite the compatibility existing between the column density 
derived in the current work and that of \citet{goldhaber_1987}, the abundance 
proposed by these authors is at least 3 times lower than ours.
This disagreement can be explained by the fact that they did not have any data 
related to the hot lines with a $T_\subscript{rot}=400$~K that we have been able 
to detect due to the high S/N of our observed spectrum.

The comparison between the abundance of \ethylene{} with respect to H$_2$
and/or the column density we derive, and the results of the chemical models
indicates that the ethylene formation process in \irc{} is still not well 
understood.
The \ethylene{} abundances derived by \citet{doty_1998} and \citet{millar_2000} 
are between one and two orders of magnitude lower than the observational 
results (Table~\ref{tab:table3}).
Additionally, these models suggest that ethylene is formed somewhere in the
outer envelope while our observations reveal that it happens much closer to the
central star.
The differences between the model of the envelope used for these chemical 
kinetics calculations and the one adopted here (e.g., our mass-loss rate is 
lower by a factor of $\simeq 2-3$) is unlikely to be the reason of the 
discrepancy concerning \ethylene. 
To our knowledge, no thermochemical equilibrium model of the surroundings 
of the star has provided predictions on the abundance of \ethylene{} in \irc. 
We have therefore carried out TE calculations for IRC+10216 to check if the 
general predictions of \citet{cherchneff_1992} can be improved for the 
particular case of \irc, fitting better our observational results.

The chemical model we have adopted is the same that \citet{fonfria_2014}
used to explain the abundances of H$^{13}$CN, SiO, and SiC$_2$ in the inner
layers of \irc{} derived from high spatial resolution observations.
The resulting \ethylene{} abundance profile is compared with the abundance
we derive from the observations in Figure~\ref{fig:f3}.
Unlike what happens for H$^{13}$CN, SiO, and SiC$_2$, that reach maximum 
abundances within $\simeq 3-4\rstar$ from the star, the predicted peak 
abundance of \ethylene{} occurs somewhat farther from the star, at 
$\simeq 9\rstar$, in very good agreement with our observational results.
In addition, this model improves the general results by \citet{cherchneff_1992}
indicating that it works reasonably well for abundant molecules (e.g., HCN or 
SiC$_2$) and for molecules such as \ethylene{} with an abundance as low as 
$\sim 10^{-8}-10^{-7}$.

On the other hand, the predicted peak abundance is more than 4 times higher 
than the value we derive from observations and the chemical model indicates 
that ethylene should vanish in a few stellar radii, something that does not 
happen according to our observations. 
This latter fact can be explained in terms of a freeze out of the TE abundance 
of \ethylene{} at radii beyond $\simeq 9\rstar$ as a consequence of the slow 
down of chemical reactions induced by the decrease in density and temperature. 
It still remains to be demonstrated whether chemical kinetics can actually be 
at work at distances from the star as large as $\simeq 9\rstar$ to drive the 
abundance of \ethylene{} to values of the order of $10^{-7}$ relative to H$_2$. 
Although observations of ethylene indicate that TE does indeed hold up to 
$\simeq 9\rstar$, a detailed chemical kinetics model dealing with the innermost 
circumstellar regions of \irc{} would be needed to shed light on this point 
from the theoretical point of view.

\subsection{Vibrational and rotational temperatures}

The vibrational temperature controls the number of molecules in the upper
vibrational state, having a direct effect on the overall emission of the
whole fundamental band $\nu_7$ and on the absorption component of the 
corresponding hot and combination bands \citep*[e.g.,][]{fonfria_2015}.
Hence, the absence of an obvious emission component in the observed lines and 
of lines of hot and combination bands above the detection limit prevents us 
from estimating this temperature.
The inner limit of the region of the envelope where \ethylene{} is formed is 
too far from the star to allow for the development of significant emission 
components and high excitation lines.
In this situation, the best choices are 
(1) to assume that \ethylene{} is under vibrational LTE or 
(2) to adopt the vibrational temperature profile of other molecules.
It was shown that molecules such as HCN, C$_2$H$_2$, and SiS 
\citep{fonfria_2008,fonfria_2015} are vibrationally out of LTE away from 
several stellar radii from the stellar photosphere so we expect this to happen 
to \ethylene{} as well.
This is supported by the fact that the collisional rate coefficients involved
in ro-vibrational transitions are expected to be several orders of magnitude
smaller than in pure rotational transitions
\citep*[e.g.,][]{tobola_2008}.

Hence, a realistic approach would be to adopt the vibrational temperature of 
the IR active C$_2$H$_2$ fundamental band $\nu_5$.
This is a reasonable choice since the adoption of a vibrational temperature 
only 10\% higher produces lines with a small though significant emission 
component, unobserved in our data.
The inaccuracy associated with this choice has implications on the 
uncertainties of the abundance, which are larger than showed in 
Table~\ref{tab:table2}.
A variation of 10\% of the vibrational temperature at 5 or 20\rstar{} modifies 
the abundance in $\simeq 2$ and 5\%, respectively, while the same relative 
change of the exponent $\gamma_\subscript{vib}$ results in a negligible variation 
of 0.5\% of the abundance.
Hence, an accurate determination of the \ethylene{} abundance would require
a better estimation of the vibrational temperature mainly between 10 and 
20\rstar. 

\begin{figure}
\centering
\includegraphics[width=0.475\textwidth]{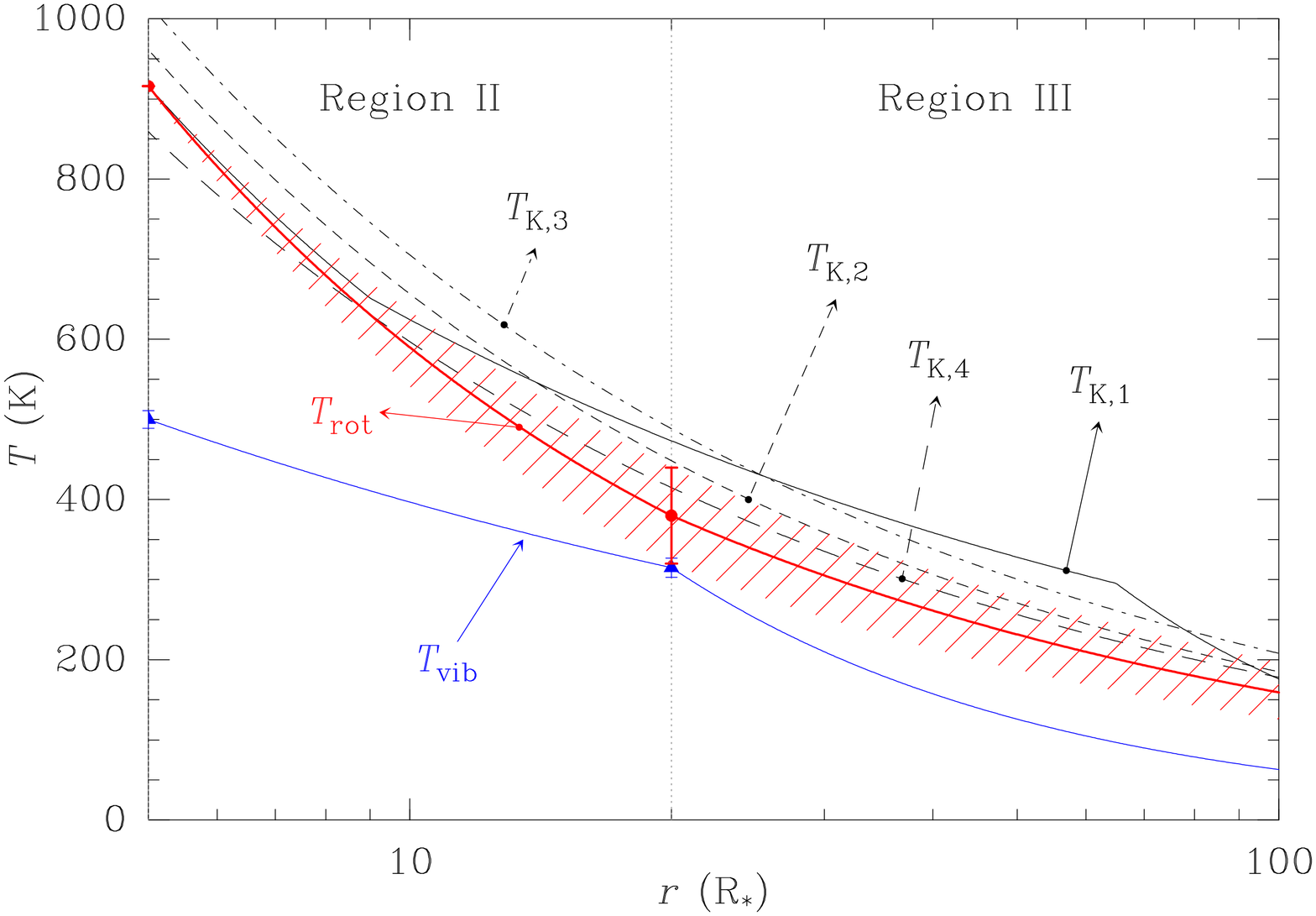}
\caption{
Kinetic, vibrational, and rotational temperatures throughout the region of the 
envelope where the observed \ethylene{} lines seem to be formed.
The kinetic temperatures $T_{\subscript{K},i}$ (black) were taken from 
(1) \citet{debeck_2012}, 
(2) \citet{agundez_2012}, 
(3) \citet{schoier_2007}, and 
(4) \citet{boyle_1994}.
The vibrational temperature $T_\subscript{vib}$ (blue) for the \ethylene{} 
$\nu_7$ band was assumed to be equal to the vibrational temperature for the
$\nu_5$ band of C$_2$H$_2$ \citep{fonfria_2008}.
The rotational temperature $T_\subscript{rot}$ (red) has been derived from
our analysis.
The $1\sigma$ error interval of $T_\subscript{rot}$ has been plotted as a striped 
region.}
\label{fig:f4}
\end{figure}

The rotational temperature, $T_\subscript{rot}$, was fixed to the kinetic one, 
$T_\subscript{K}$, at 5\rstar, where the \ethylene{} abundance probably is very 
low and we have no reliable information about the excitation.
However, the rotational temperature at 20\rstar{} can be properly determined
from the fits to the observations adopting a value of $\simeq 380$~K with an 
uncertainty of $\simeq 15\%$ (Table~\ref{tab:table2}).
Beyond this position, our best fit results in a power-law dependence 
$\propto r^{-0.54}$.
A decrease of 50\% in the rotational temperature at 5\rstar{} only introduces 
changes in the synthetic spectrum up to the noise level ($\sigma=0.05\%$) so 
that the rotational temperature at 20\rstar{} remains unaffected.
The comparison between the rotational and the kinetic temperatures by 
\citet{debeck_2012} suggests that \ethylene{} could be rotationally out of LTE 
(Figure~\ref{fig:f4}).
However, if we compare $T_\subscript{rot}$ with \tk{} taken from other works
\citep{boyle_1994,schoier_2007,agundez_2012}, the $1\sigma$ error intervals of 
$T_\subscript{rot}$ and the actual \tk{} are overlapped.
Thus, the deviation of the rotational temperature to the kinetic one
is not significant and \ethylene{} can be assumed rotationally under LTE, as it 
is expected for a symmetric molecule with no permanent dipole moment.

This rotational temperature, that we have concluded to be very similar to the
actual kinetic temperature, allows us to refine the zones of the envelope where 
the bulk absorption of the warm and hot lines found in the ro-vibrational 
diagram (Fig.~\ref{fig:f2}) arise.
The warm lines, with an average rotational temperature of $105\pm 30$~K, are 
mostly formed between 135 and 400\rstar{} while the hot lines, with an average 
rotational temperature of $400\pm 80$~K, are formed between 14 and 28\rstar.

\subsection{Ethylene and dust}

Although the abundance profile in Region II derived from Scenario $2A$ is
too rough to get information about the \ethylene{} depletion onto the dust 
grains, Scenario $2B$ suggests that it might happen at 20\rstar{} with at most 
25\% of the abundance with respect to H$_2$ in the outer shells of Region II 
($\simeq 1.1\times 10^{-7}$).
This means that the density of \ethylene{} molecules in these shells that could
condense is up to 0.9~cm$^{-3}$, implying an upper limit for the column density 
along 1\rstar{} of $\simeq 3.3\times 10^{13}$~cm$^{-2}$.
Thus, a 1\rstar{} width shell with a radius of 20\rstar{} contains up to 
$2.3\times 10^{44}$ \ethylene{} molecules, i.e., $1.1\times 10^{22}$~g of 
ethylene.
Assuming that the total mass ejected from the star in form of dust is
$\simeq 2.5\times 10^{-8}$~M$_\odot$~yr$^{-1}$ \citep{debeck_2012,decin_2015},
there is about $4.6\times 10^{25}$~g of dust in the shell defined above.
Hence, the possible contribution of \ethylene{} to the dust mass around 
20\rstar, where the gas seems to be accelerated from 11 to 14.5~\kms,
would be $\lesssim 0.025\%$ corresponding to an increment of the size of the 
dust grains $\lesssim 0.02$\%.
This negligible growth cannot affect the gas and dust acceleration but the 
\ethylene{} condensation could enrich the chemistry on the surface of the 
grains.

\section{Summary}
\label{sec:summary}

High resolution observations ($R\simeq 100,000$) of the \ethylene{} band 
$\nu_7$ found in the $\simeq 10.5~\mu$m spectrum of \irc{} have been analyzed.
Eighty \ethylene{} lines were found in the observations above the $3\sigma$ 
($\simeq 0.15\%$ of the continuum) detection limit. 
Of these lines about 35\% are unblended or partially blended allowing for a 
detailed analysis of their profiles with a rotational diagram and the code 
developed by \citet{fonfria_2008}.

From these analyses, we conclude that:
\begin{itemize}
\item The \ethylene{} lines can be divided into two groups with significantly 
different rotational temperatures: a hot group with an rotational temperature 
of $\simeq 400$~K composed of lines located between $\simeq 14$ and 28\rstar{}
and a warm group with an rotational temperature of $\simeq 105$~K composed of 
lines mostly produced between $\simeq 135$ and 400\rstar.
No evidence of \ethylene{} closer to the star than $\simeq 10$\rstar{} was 
found.
\item The \ethylene{} abundance profile with respect to H$_2$ is compatible 
with an average value of $6.9\times 10^{-8}$ between 5 and 20\rstar{} and
a terminal abundance of $8.2\times 10^{-8}$ beyond 20\rstar.
This last value is at least 4 times larger than the chemical models predict.
The terminal abundance seems not to depend on the chosen profile but the 
abundance closer to the star is probably more complex that we assumed, 
involving its growth between 5 and 10\rstar.
The total \ethylene{} column density ranges from 4.2 to $7.5\times 10^{15}$~cm$^{-2}$.
\item The rotational temperature derived from the fits to the observed lines
is equal to the kinetic temperature at 5\rstar{} ($\simeq 915$~K) and
$\simeq 380$~K at 20\rstar, below the kinetic temperature ($\simeq 475$~K).
At larger distances from the star the rotational temperature follows the 
power-law $\propto r^{-0.54}$.
\ethylene{} can be considered rotationally under LTE throughout the region of 
the envelope traced with our observations.
\item The vibrational temperature of the \ethylene{} band $\nu_7$ is unknown 
but it could be assumed to be similar to that of the C$_2$H$_2$ band $\nu_5$.
In all cases, the good fits achieved with this temperature suggests that 
\ethylene{} is vibrationally out of LTE.
\item A fraction of the gas-phase \ethylene{} could condense onto the dust 
grains around $\simeq 20\rstar$.
The dust grains barely grow due to this depletion and no significant influence 
on the gas acceleration is expected to happen but the adsorbed \ethylene{} 
molecules could affect the chemistry evolution in the surface.
\end{itemize}

\acknowledgments

We thank the anonymous referee for his/her help in the correction and
improvement of this manuscript.
Development of TEXES was supported by grants from the NSF and USRA.
MJR and others thank IRTF, which is operated by the University of Hawaii under 
Cooperative Agreement NCC 5-538 with the National Aeronautics and Space 
Administration, Office of Space Science, Planetary Astronomy Program.
The National Optical Astronomy Observatory is operated by the Association 
of Universities for Research in Astronomy (AURA) under cooperative agreement 
with the National Science Foundation.
JC and JPF thank the Spanish MINECO/MICINN for funding support through grants
AYA2009-07304, AYA2012-32032, the ASTROMOL Consolider project CSD2009-00038 and
the European Research Council (ERC Grant 610256: NANOCOSMOS).

\facility{IRTF(TEXES)}


\begin{thebibliography}{}
\bibitem[Ag\'undez \& Cernicharo(2006)]{agundez_2006}           Ag\'undez M. \& Cernicharo J., 2006, \apj, 650, 374
\bibitem[Ag\'undez et al.(2010)]{agundez_2010}                  Ag\'undez M., Cernicharo J. \& Gu\'elin M., 2010, \apj, 724, L133
\bibitem[Ag\'undez et al.(2012)]{agundez_2012}                  Ag\'undez M., Fonfr\'ia J. P., Cernicharo J., Kahane C., Daniel F., \& Gu\'elin M., 2012, \aap, 543, A48
\bibitem[Ag\'undez et al.(2014)]{agundez_2014}                  Ag\'undez M., Cernicharo J. \& Gu\'elin M., 2014, \aap, 570, A45
\bibitem[Bernath et al.(1989)]{bernath_1989}                    Bernath P. F., Hinkle K. W. \& Keady J. J., 1989, \science, 244, 562
\bibitem[Betz(1981)]{betz_1981}                                 Betz A. L., 1981, \apj, 244, L103
\bibitem[Blass et al.(2001)]{blass_2001}                        Blass W. E. et al., 2001, \jqsrt, 71, 47
\bibitem[Boyle et al.(1994)]{boyle_1994}                        Boyle R. J., Keady J. J., Jennings D. E., Hirsch K. L. \& Wiedemann G. R., 1994, \apj, 420, 863
\bibitem[Cernicharo et al.(1987)]{cernicharo_1987}              Cernicharo J. \& Gu\'elin M., 1987, \aap, 183, L10
\bibitem[Cernicharo et al.(1989)]{cernicharo_1989}              Cernicharo J., Gottlieb C. A., Gu\'elin M., Thaddeus P. \& Vrtilek J. M., 1989, \apj, 341, L25
\bibitem[Cernicharo et al.(1999)]{cernicharo_1999}              Cernicharo J., Yamamura I., Gonz\'alez-Alfonso E., de Jong T., Heras A., Escribano R., \& Ortigoso J., 1999, \apj, 526, L41
\bibitem[Cernicharo et al.(2000)]{cernicharo_2000}              Cernicharo J., Gu\'elin M., \& Kahane C., 2000, \aaps, 142, 181
\bibitem[Cernicharo(2004)]{cernicharo_2004}                     Cernicharo J., 2004, \apj, 608, L41
\bibitem[Cernicharo et al.(2010)]{cernicharo_2010}              Cernicharo J. et al., 2010, \aap, 521, L8
\bibitem[Cernicharo et al.(2013)]{cernicharo_2013}              Cernicharo J., Daniel F., Castro-Carrizo A., Ag\'undez M., Marcelino N., Joblin C., Goicoechea J. R. \& Gu\'elin M., 2013, \apj, 778, L25
\bibitem[Cernicharo et al.(2015a)]{cernicharo_2015a}            Cernicharo J., Marcelino N., Ag\'undez M. \& Gu\'elin M., 2015, \aap, 575, A91
\bibitem[Cernicharo et al.(2015b)]{cernicharo_2015b}            Cernicharo J. et al., 2015b, \apj, 806, L3
\bibitem[Cernicharo et al.(2016)]{cernicharo_2016}              Cernicharo J., Castro-Carrizo A., Quintana-Lacaci G., Ag\'undez M., Marcelino N., Velilla Prieto L., Fonfr\'ia J. P. \& Gu\'elin M., 2016, \textit{submitted to the ApJ}
\bibitem[Cherchneff \& Barker(1992)]{cherchneff_1992}           Cherchneff I. \& Barker J. R., 1992, \apj, 394, 703
\bibitem[Cherchneff(2006)]{cherchneff_2006}                     Cherchneff I., 2006, \aap, 456, 1001
\bibitem[Contreras et al.(2011)]{contreras_2011}                Contreras C. S., Ricketts C. L. \& Salama F., 2011, EAS Publication Series, 46, 201
\bibitem[De Beck et al.(2012)]{debeck_2012}                     De Beck et al., 2012, \aap, 539, A108
\bibitem[Decin et al.(2010a)]{decin_2010a}                      Decin L. et al., 2010a, \aap, 518, L143
\bibitem[Decin et al.(2010b)]{decin_2010b}                      Decin L. et al., 2010b, \nature, 467, 64
\bibitem[Decin et al.(2015)]{decin_2015}                        Decin L., Richards A. M. S., Neufeld D., Steffen W., Melnick G. \& Lombaert R., 2015, \aap, 574, 5
\bibitem[Doty \& Leung(1998)]{doty_1998}                        Doty S. D. \& Leung C. M., 1998, \apj, 502, 898
\bibitem[Encrenaz et al.(1975)]{encrenaz_1975}                  Encrenaz T., Combes M., Zeau Y., Vapillon L. \& Berezne J., 1975, \aap, 42, 355
\bibitem[Fonfr\'ia et al.(2008)]{fonfria_2008}                  Fonfr\'ia, J. P., Cernicharo, J., Richter, M. J., \& Lacy, J. H., 2008, \apj, 673, 445
\bibitem[Fonfr\'ia et al.(2011)]{fonfria_2011}                  Fonfr\'ia, J. P., Cernicharo, J., Richter, M. J., \& Lacy, J. H., 2011, \apj, 728, 43
\bibitem[Fonfr\'ia et al.(2014)]{fonfria_2014}                  Fonfr\'ia J. P., Fern\'andez-L\'opez M., Ag\'undez M., S\'anchez-Contreras C., Curiel S., \& Cernicharo J., 2014, \mnras, 445, 3289
\bibitem[Fonfr\'ia et al.(2015)]{fonfria_2015}                  Fonfr\'ia J. P., Cernicharo J., Richter M. J., Fern\'andez-L\'opez M., Velilla Prieto L. \& Lacy J. H., 2015, 2015, \mnras, 453, 439
\bibitem[Glassgold(1996)]{glassgold_1996}                       Glassgold A. E., 1996, \araa, 34, 241
\bibitem[Gray et al.(1977)]{gray_1977}                          Gray D. L., Robiette A. G., \& Johns, J. W. C. 1977, \molphys, 34, 1437
\bibitem[Groenewegen et al.(2012)]{groenewegen_2012}            Groenewegen M. A. T. et al., 2012, \aap, 543, L8
\bibitem[Goldhaber et al.(1987)]{goldhaber_1987}                Goldhaber D. M., Betz A. L. \& Ottusch J. J., 1987, \apj, 314, 356
\bibitem[Gu\'elin \& Cernicharo(1991)]{guelin_1991}             Gu\'elin M. \& Cernicharo J., 1991, \aap, 244, L21
\bibitem[Gu\'elin et al.(1997)]{guelin_1997}                    Gu\'elin M., Lucas R. \& Neri R., 1997, IAU Symposium 170, CO: Twenty-five years of millimeter spectroscopy, eds. W. B. Latter, S. Radford, P. R. Jewell, J. G. Mangum \& J. Bally (Kluwer, Dordrecht, The Netherlands), pp. 359-366
\bibitem[He et al.(2008)]{he_2008}                              He J. H., Dinh-V-Trung, Kwok S., M\"uller H. S. P., Zhang Y., Hasegawa T., Peng T. C., \& Huang Y. C., 2008, \apj, 177, 275
\bibitem[Hinkle et al.(1988)]{hinkle_1988}                      Hinkle K. H., Keady J. J. \& Bernath P. F., 1988, \science, 241, 1319
\bibitem[Hinkle et al.(2008)]{hinkle_2008}                      Hinkle K. H., Wallace L., Richter M. J., \& Cernicharo J., 2008, Proceedings IAU Symposium, 251, 161
\bibitem[Kawaguchi et al.(1995)]{kawaguchi_1995}                Kawaguchi K., Kasai Y., Ishikawa S.-I., \& Kaifu N., 1995, \pasj, 47, 853
\bibitem[Keady et al.(1988)]{keady_1988}                        Keady J. J., Hall D. N. B. \& Ridgway S. T., 1988, \apj, 326, 832
\bibitem[Keady \& Ridgway(1993)]{keady_1993}                    Keady J. J. \& Ridgway S. T., 1993, \apj, 406, 199
\bibitem[Kim et al.(1985)]{kim_1985}                            Kim S. J., Caldwell J., Rivolo A. R. \& Wagener R., 1985, \icarus, 64, 233
\bibitem[Kim et al.(2015)]{kim_2015}                            Kim H., Lee H.-G., Mauron N. \& Chu Y.-H., 2015, \apj, 804, L10
\bibitem[Lacy et al.(2002)]{lacy_2002}                          Lacy J. H., Richter M. J., Greathouse T. K., Jaffe D. T., \& Zhu Q., 2002, \pasp, 114, 153
\bibitem[Lafferty et al.(2011)]{lafferty_2011}                  Lafferty W. J., Flaud J. M., \& Kwabia F., 2011, \molphys, 109, 2501
\bibitem[Lebron \& Tan(2013a)]{lebron_2013a}                    Lebron G. B. \& Tan T. L., 2013a, \ijs, 492092
\bibitem[Lebron \& Tan(2013b)]{lebron_2013b}                    Lebron G. B. \& Tan T. L., 2013b, \jms, 283, 29
\bibitem[Lebron \& Tan(2013c)]{lebron_2013c}                    Lebron G. B. \& Tan T. L., 2013c, \jms, 288, 11
\bibitem[Loro\~no-Gonz\'alez et al.(2010)]{loronogonzalez_2010} Loro\~no Gonz\'alez M. A. et al., \jqsrt, 111, 2151
\bibitem[Millar et al.(2000)]{millar_2000}                      Millar T., Herbst E. \& Bettens R. P. A., 2000, \mnras, 316, 195
\bibitem[Mutschke et al.(1999)]{mutschke_1999}                  Mutschke H., Andersen A. C., Cl\'ement D., Henning Th., \& Peiter G., 1999, \aap, 345, 187
\bibitem[Oomens et al.(1996)]{oomens_1996}                      Oomens J., Reuss J., Mellau G. Ch., Klee S., Gulaczyk I., \& Fayt A., 1996, \jms, 180, 236
\bibitem[Patel et al.(2011)]{patel_2011}                        Patel N. A. et al., 2011, \apjs, 193, 17
\bibitem[Pierre et al.(1986)]{pierre_1986}                      Pierre G., Valentin A. \& Henry L. 1986, \cjp, 64, 341
\bibitem[Quintana-Lacaci et al.(2016)]{quintana-lacaci_2016}    Quintana-Lacaci G. et al., 2016, \apj, 818, 192
\bibitem[Ridgway et al.(1988)]{ridgway_1988}                    Ridgway S. T. \& Keady J. J., 1988, \apj, 326, 843
\bibitem[Rothman et al.(2013)]{rothman_2013}                    Rothman L. S. et al., 2013, \jqsrt, 130, 4
\bibitem[Rouleau \& Martin(1991)]{rouleau_1991}                 Rouleau F. \& Martin P. G., 1991, \apj, 377, 526
\bibitem[Sartakov et al.(1997)]{sartakov_1997}                  Sartakov B. G., Oomens J., Reuss J., Fayt A., 1997, \jms, 185, 31
\bibitem[Sch\"oier et al.(2007)]{schoier_2007}                  Sch\"oier F. L., Bast J., Olofsson H. \& Lindqvist M., 2007, \aap, 473, 871
\bibitem[Schulz et al.(1999)]{schulz_1999}                      Schulz B., Encrenaz Th., B\'ezard B., Romani P. N., Lellouch E. \& Atreya S. K., 1999, \aap, 350, L13
\bibitem[Tallon-Bosc et al.(2007)]{tallon-bosc_2007}            Tallon-Bosc I., Tallon M., Thi\'ebaut E. \& B\'echet C., 2007, \nar, 51, 697
\bibitem[Taylor(1997)]{taylor_1997}                             Taylor J. R. in ``An introduction to error analysis: the study of uncertainties in physical measurements'', University Science Books, Sausalito CA, 2nd ed., 1997, ISBN: 0-935702-42-3 
\bibitem[Tobo{\l}a et al.(2008)]{tobola_2008}                   Tobo{\l}a R., Lique F., K{\l}os J. \& Cha{\l}asi\'nski G., 2008, \jpb, 41, 155702
\bibitem[Ulenikov et al.(2013)]{ulenikov_2013}                  Ulenikov O. N., Gromova O. V., Aslapovskaya Yu. S., \& Horneman V.-M., 2013, \jqsrt, 118, 14
\bibitem[Willacy \& Cherchneff(1998)]{willacy_1998}             Willacy K. \& Cherchneff I., 1998, \aap, 330, 676
\bibitem[Willaert et al.(2006)]{willaert_2006}                  Willaert F., Demaison J., Margules L., M\"ader H., Spahn H., Giesen T., \& Fayt A., 2006, \molphys, 104, 273
\bibitem[Woods et al.(2003)]{woods_2003}                        Woods P. M., Millar T. J., Herbst E. \& Zijlstra A. A., 2003, \aap, 402, 189
\end{thebibliography}
\end{document}